\documentclass[11pt]{article}

\usepackage{graphicx}
\usepackage{amsmath}
\usepackage{amssymb}
\usepackage{enumerate}
\usepackage{xcolor}
\usepackage{bm}
\usepackage[T1]{fontenc}
\usepackage{float}
\usepackage{soul} % \st{}
\usepackage{multirow}
\usepackage{comment}
\usepackage{moreverb,url}
\usepackage{setspace}

\usepackage{booktabs}
\usepackage{tabularx}

\newcommand\clearrow{\global\let\rowmac\relax}
\clearrow

\usepackage{natbib}

\usepackage{amsthm,amssymb,amsmath}

\def \bs { \boldsymbol  s}
\def \bY { \boldsymbol Y }
\def \by { \boldsymbol y }
\def \bt { \boldsymbol t }

\def \bX { \boldsymbol X }
\def \bZ { \boldsymbol Z }
\def \bV { \boldsymbol V }

\def \bb { \boldsymbol b }
\def \bB { \boldsymbol B }

\def \bD { \boldsymbol D }

\def \bu { \boldsymbol u }
\def \bI { \boldsymbol I }

\def \bbeta { \boldsymbol \beta }
\def \balpha { \boldsymbol \alpha }
\def \bgamma { \boldsymbol \gamma }

\def \bTheta { \boldsymbol \Theta }

\def \bSigma { \boldsymbol \Sigma }

\def \bepsilon { \boldsymbol \epsilon }

\def \bxi { \boldsymbol \xi }

\def \tT { \tilde{T} }
\def \tD { \tilde{D} }

\def\bSig\mathbf{\Sigma}

\title{\textbf{Backward Joint Model for the Joint Dynamic Prediction of Time-to-Event and Longitudinal Data: Basic Formulation and New Developments}}

\date{}
\begin{document}

\maketitle

\begin{center}
\author{Wenhao Li$^{1}$,   \and
        Shikun Wang$^{2}$, \and 
        Zhe Yin$^{1}$,     \and
        Brad C. Astor$^{3}$, \and 
        Wei Yang$^{4}$,     \\ 
        Tom H. Greene$^{5}$ and \and 
        Liang Li$^{1*}$ 
}
\end{center}

\begin{center}
$^{1}$ Department of Biostatistics, The University of Texas MD Anderson Cancer Center, Houston, TX, USA;  \\
           $^{2}$ Department of Biostatistics, Columbia University, New York, NY, USA; \\
           $^{3}$ Department of Population Health Sciences, The University of Wisconsin, Madison, WI, USA;  \\ 
           $^{4}$ Department of Biostatistics, Epidemiology and Informatics, The University of Pennsylvania, Philadelphia, PA, USA;   
           \\
           $^{6}$ Department of Population Health Sciences, The University of Utah, Salt Lake City, UT, USA; \\ 
           *Liang Li: LLi15@mdanderson.org
\end{center}

\begin{abstract}
	Dynamic prediction of future clinical outcomes based on longitudinally measured predictors plays a crucial role in disease management and patient counseling, particularly when conventional static models are inadequate. Joint modeling of longitudinal and time-to-event data provides a useful framework for addressing this challenge. In this paper, we present a comprehensive development of the recently proposed backward joint model (BJM; \citealt{Shen2021}), which factorizes the likelihood into the distribution of time-to-event data and the conditional distribution of longitudinal data given the event time. This structure facilitates computation and is well-suited for multivariate longitudinal data. We introduce several novel developments to the BJM, including the extrapolation and two-part specifications, as well as the incorporation of competing risks. We also address an important yet underexplored problem in the literature: predicting future longitudinal trajectories conditional on predicted event times. Additionally, we explore the connection between BJM and existing joint modeling approaches. All these extensions preserve the computational advantages of the basic BJM formulation, including one-dimensional numerical integration, convex optimization via the EM algorithm, and a quick procedure for consistent estimation using standard software. We evaluate the method’s performance through simulation studies and illustrate its utility in a chronic kidney disease application.\\
\end{abstract}

\noindent%
\vfill

\section{Introduction}
\label{sec:intro}

\subsection{\textbf{Motivating applications}}
\label{subsec:motivating.applications}

Chronic kidney disease (CKD) is a major public health burden. It is characterized by a gradual and often irreversible deterioration of kidney function over years, and is progressively associated with many comorbidities and adverse outcomes, including diabetes, hypertension, and cardiovascular disease \citep{appel2023estimated}. The stages of CKD are defined based on the kidney function biomarker, glomerular filtration rate (GFR, in mL/min/1.73 m$^2$). Stages 1, 2, 3A, 3B, 4 and 5 correspond to GFR in $\geqslant 90$, $60-89$, $45-59$, $30-44$, $15-29$, and $< 15$ mL/min/1.73 m$^2$, respectively. At the low end of stage 4 or stage 5, patients need to be prepared for entering End-Stage Renal Disease (ESRD), when either dialysis or kidney transplant is required. Once a CKD patient enters ESRD, the GFR is not applicable. Hence, we call the ESRD a ``terminal'' event. Death is another terminal event. 

Predicting ESRD is crucial for research, clinical management, and patient counseling. An effective prediction model should address several key statistical challenges. First, it must adapt to a dynamic at-risk population and time-varying associations between predictors and outcomes \citep{Li2016}. Changes in the at-risk population arise from aging and systematic differences between early and late progressors, while predictor relevance may shift over the disease course. Second, the model should incorporate patients’ clinical history, not just their current health status. Third, patients differ in their concerns: those in early disease stages may care more about long-term GFR trajectories assuming survival, rather than immediate ESRD risk. Finally, the model must account for the competing risk of death to avoid bias and ensure valid interpretation \citep{Li2019toolbox}.

\subsection{\textbf{Overview of dynamic prediction (DP)}}
\label{subsec:overview.DP}

The statistical challenges above can be addressed within the framework of dynamic prediction \citep{Taylor2005, Yu2008, Rizopoulos2011}, which can be formulated mathematically as follows. Suppose the training dataset includes $n$ subjects indexed by $i = 1, 2, \dots, n$. Let $\tT_i$ be the time to a terminal event and $C_i$ the censoring time. The observed time is $T_i = \min(\tT_i, C_i)$, with censoring indicator $\delta_i = \bm{I}(\tT_i < C_i)$. Let $\bV_i$ denote the vector of baseline covariates. Assume $M$ continuous biomarkers are measured longitudinally, indexed by $m = 1, 2, \dots, M$. For subject $i$, let $\bY_{mi}$ be the repeated measurements of the $m$-th biomarker, observed at the follow-up visit times $\bt_{mi}$, with $n_{mi}$ measurements. Measurement times may be irregular and differ across biomarkers and subjects. Let $\bY_i = (\bY_{1i}^T, \dots, \bY_{Mi}^T)^T$ and $\bt_i = (\bt_{1i}^T, \dots, \bt_{Mi}^T)^T$ denote the stacked vectors of all longitudinal biomarker values and their corresponding times. We use two conventional assumptions in the joint modeling literature: non-informative observation times, which requires the distribution of $\bt_i$ in $(0, T_i)$ to be independent of the rest of the data, and independent censoring, which requires $(\bY_i, \tT_i) \perp C_i | \bV_i $. 

We use the subscript $o$ to index a generic subject from the validation data, with a distribution identical to that of the training data, for whom the prediction will be made at time $s$. Let $\overline{\bm{Y}_o(s)}$ represent the longitudinal biomarker history prior to $s$ (i.e., all the biomarker measurements before $s$). $\Delta$ denotes a pre-specified prediction horizon. The DP literature usually studies the following estimand, which pertains to the risk of a future clinical event:
\begin{equation}
\begin{aligned}
	P(s < \tT_o \leqslant s + \Delta | \overline{\bY_o(s)}, T_o > s, \bV_o) = \dfrac{P( \overline{\bm{Y}_o(s)},  s<\tilde{T}_o \leqslant s +\Delta |\bV_o) }{P( \overline{\bm{Y}_o(s)}, \tT_o > s| \bV_o)}  . 
    \label{DP_risk}
\end{aligned}
\end{equation} 
We refer to this equation as the \textit{dynamic prediction formula}. The left-hand side is the probability of the event in the next $\Delta$ years (or other time units), given that the subject is still at-risk at the prediction time $s$ ($T_o > s$) and given the observed history ($\bV_o$, $\overline{\bY_o(s)}$). This estimand addresses the first two statistical challenges in Section \ref{subsec:motivating.applications}. The $T_o$ on the left hand side is eliminated on the right-hand side due to the independent censoring assumption. 

The left of (\ref{DP_risk}) suggests the landmark modeling approach to DP, i.e., fitting a regression model of residual lifetime $\tT_i(s) = \tT_i - s$ given the covariates from $\overline{\bY_i(s)}$ and $\bV_i$, among the at-risk subjects with $T_i > s$ \citep{Zheng2005, vanHouwelingen2011, Li2016, Wu2020}. The right hand suggests the joint modeling approach because the prediction can be calculated from $(\bY, \tT) | \bV $, a model for the joint distribution of longitudinal and survival data \citep{Rizopoulos2012, Elashoff2016, Andrinopoulou2021}. This formula encapsulates the two mainstream approaches to DP in a single equation.  

In a longitudinal setting such as  Section~\ref{subsec:motivating.applications}, the DP model is, in theory, more accurate than the commonly used static prediction model (SPM) because it is adaptive to the time-varying at-risk population and longitudinal history. The SPM is trained by regressing the survival outcome on baseline predictors (including baseline $\bY$ as part of $\bV$), then applied at the time $s$ using predictor values observed at that time \citep{Parast2019, Yao2023}. Formally, let $\tilde{T}$ denote the event time and $\overline{\bY(\tT)} = \{ \bY(t); t \in (0, \tT) \}$ denote the history of all predictors up to $\tT$. The conditional distribution of survival data given the baseline predictor is $f_1(\tT | \bY(0); \balpha)$ with parameter $\balpha$. The joint distribution of longitudinal and survival data is $f_2(\tilde{T}, \overline{\bY(\tT)}; \bbeta)$ with parameter $\bbeta$. For an at-risk subject at time $s$, the static prediction is 
	\begin{equation}
		m_1( \bY(s); \balpha ) = P(\tT - s > \Delta | \tT > s, \bY(s); \balpha ) = \int_{\Delta}^{\infty} f_1( t | \bY(s) ; \balpha ) dt . \notag   
	\end{equation}
	The dynamic prediction is
	\begin{equation}
		m_2(\overline{\bY(s)}; \bbeta) = P(\tT - s > \Delta | \tT > s, \overline{\bY(s)}; \bbeta )  .   \notag 
	\end{equation}
	Then, among subjects in the risk set $\tT > s$, the DP model has equal or better prediction accuracy than the SPM in terms of the Brier score, i.e., 
	\begin{equation}
    E\left\{ L_s\left( m_2(\overline{\mathbf{Y}(s)}; \boldsymbol{\beta}) \right) \mid \tT > s \right\} \leqslant E\left\{ L_s\left( m_1(\mathbf{Y}(s); \boldsymbol{\alpha}) \right) \mid \tT > s \right\} , 
    \label{eq:SPM.DPM}
    \end{equation}
where $L_{s}(P) = ( 1\{ \tT - s > \Delta \} - P)^2$ is the squared error loss for predicted probability $P$ at prediction time $s$. The proof is in the Online Appendix. In practice, the relative performance of SPM and DP may be influenced by model misspecification, finite-sample variability, and how effectively clinical history is incorporated. Nonetheless, this theoretical result provides support for the use of DP. \citet{Li2023} showed that both landmark and joint modeling approaches are valuable and merit further investigation.

\subsection{\textbf{Dynamic prediction of future longitudinal data}}
\label{subsec:predict.longitudinal}

We now address the third statistical challenge in Section \ref{subsec:motivating.applications}. Let $P(.)$ denote probability or probability density, when there is no ambiguity. We are interested in predicting the distribution of $\bY_o(t)$, the generic subject's longitudinal biomarker at a future time $t > s$:     
\begin{equation}
	P(\bY_o(t) = \by | \overline{\bm{Y}_o(s)}, T_o > s , \bV_o, \tT_o > t ) = \frac{P(\bY_o(t)  = \by,  \overline{\bm{Y}_o(s)}, \tT_o > t | \bV_o) }{P(\overline{\bm{Y}_o(s)}, \tT_o > t | \bV_o)}.
    \label{DP_eGFR}
\end{equation}
This is the probability density function of longitudinal biomarkers at a future time $t$, conditional on the subject being at risk ($T_o > s$), with observed history $\{ \overline{\bm{Y}_o(s)}, \bV_o \}$ and the absence of the future terminal event by $t$. The equality is derived from Bayes theorem and the independent censoring assumption. A key innovation of this formulation is that future biomarker values are only defined if the terminal event has not occurred. This predictive estimand has not been studied in the literature previously. Once the density in (\ref{DP_eGFR}) is estimated, summary features --- such as predicted mean, mode, and quantiles --- can all be calculated, as demonstrated in our analysis. The future time $t$ may vary to yield a subject-specific predicted trajectory. However, this trajectory should not be viewed as a simple extrapolation from past biomarker trends, since it incorporates the risk of the terminal event, which varies with $t$. 

Unlike the dynamic prediction formula (\ref{DP_risk}), the left-hand side of (\ref{DP_eGFR}) does not suggest a straightforward landmark modeling approach for $\bY_o$ because of the conditioning on $\tT_o > t$. Nonetheless, the right-hand side can still be calculated from joint modeling, as illustrated later. 

% Prior work on predicting longitudinal trajectories \citep{Taylor2005, Yu2008, Rizopoulos2012} does not condition on the absence of future terminal events, and thus applies only when the outcome is non-terminal (i.e., when $\bY$ remains defined after $\tT$). In contrast, the estimand in (\ref{DP_eGFR}) is specifically tailored for settings involving terminal events, such as Section \ref{subsec:motivating.applications}.

\subsection{\textbf{Outline of this paper}}
\label{subsec:outline}

We refer to the computation of both (\ref{DP_risk}) and (\ref{DP_eGFR}) for a generic subject in the validation data as a \emph{joint dynamic prediction} problem. This is the primary analytical goal of this paper. Since both estimands are functions of the conditional joint distribution $P(\bY_o, \tT_o | \bV_o)$, we concentrate on estimating this distribution using the training data. While various models can be employed, such as the widely used shared random effects joint model \citep{Rizopoulos2012, Elashoff2016}, this paper focuses on the backward joint model (BJM), initially proposed by \citet{Shen2021} for the dynamic prediction in (\ref{DP_risk}). The BJM offers advantages in computational speed and stability, flexible specification, straightforward diagnostics, and easy implementation using standard software. We extend the BJM into a comprehensive framework for joint dynamic prediction, incorporating several new developments: a dual formulation (extrapolation and two-part models; Section~\ref{subsec:twoBJM}), flexible submodel choices (Sections~\ref{subsec:survival.longitudinal.submodels}), links to other joint models (Sections~\ref{subsec:literature} and \ref{subsec:connection}), modeling assumptions assessment (Section \ref{subsec:censoring}), and an extension to competing risk outcomes (Section~\ref{sec:BJM.comprisk}). Because right-censored time-to-event data are a special case of competing risks, estimation (Section~\ref{sec:training}), data analysis (Section~\ref{sec:data_application}), and simulation (Section~\ref{sec:Simulation}) are all presented within this broader context. Summary and discussions are in Section \ref{sec:Discussion}. Additional supporting materials are provided in the Online Appendix.

\section{The Backward Joint Model (BJM): Basic Formulation}
\label{sec:BJM.basic}

\subsection{\textbf{The likelihood factorization}}
\label{subsec:twoBJM}

The BJM factorizes the joint distribution of interest into two conditionals,
\begin{equation}
P(\bY_o, \tT_o \mid \bV_o) = P(\tT_o \mid \bV_o) P(\bY_o \mid \tT_o, \bV_o),
\label{eq:factorize}
\end{equation}
with each conditional distribution estimated by a sub-model separately (Section \ref{subsec:survival.longitudinal.submodels}). Since $\bY_o$ precedes $\tT_o$, this is referred to as the ``backward'' joint model, a term introduced by \citet{vanHouwelingen2011}. Related methods and literature are discussed in Section~\ref{subsec:connection}.  

Estimands (\ref{DP_risk}) and (\ref{DP_eGFR}) both involve this joint distribution in the denominator: 
\begin{equation}
P( \overline{\bm{Y}_o(s)}, \tT_o > t \mid \bV_o ) = \int_t^{\infty} P( \overline{\bm{Y}_o(s)} \mid \tT_o = u, \bV_o ) \, P( \tT_o = u \mid \bV_o ) \, du, \quad t \geqslant s. \notag     
\end{equation}
In typical biomedical studies, both $\tT_o$ and $C_o$ have finite support. We consider two study designs. In the first, there exists a time $\tau_{\text{max}}$ such that $P(C_o > \tau_{\text{max}} \mid \bV_o) > 0$ and $P(\tT_o > \tau_{\text{max}} \mid \bV_o) = 0$. Under this setting, both conditional distributions in (\ref{eq:factorize}) are identifiable for all $u$ in the support of $\tT_o$, and a sub-model is required for each. This corresponds to a design where follow-up typically exceeds the survival time of study participants.

In the second scenario, there exists a time $\tau_{\text{max}}$ such that $P(\tT_o > \tau_{\text{max}} \mid \bV_o) > 0$ and $P(C_o > \tau_{\text{max}} \mid \bV_o) = 0$ for at least some subjects. In this case, the conditional distributions $P( \overline{\bm{Y}_o(s)} \mid \tT_o = u, \bV_o )$ and $P( \tT_o = u \mid \bV_o )$ are not identifiable from the data when $u > \tau_{\text{max}}$, as follow-up is insufficient to estimate them over their entire support. To address this, we rewrite
\begin{equation}
\begin{aligned}
P( \overline{\bm{Y}_o(s)}, \tT_o > t \mid \bV_o ) = {} & \int_t^{\tau_{\text{max}}} P( \overline{\bm{Y}_o(s)} \mid \tT_o = u, \bV_o ) \, P( \tT_o = u \mid \bV_o ) \, du \\
& + P( \overline{\bm{Y}_o(s)} \mid \tT_o > \tau_{\text{max}}, \bV_o ) \, P( \tT_o > \tau_{\text{max}} \mid \bV_o ). \label{eq:integral.split}
\end{aligned}
\end{equation}
A survival sub-model can identify $P(\tT_o = u \mid \bV_o)$ for $t < u \leqslant \tau_{\text{max}}$ and $P(\tT_o > \tau_{\text{max}} \mid \bV_o)$. For the longitudinal component, two sub-models are needed: one for $P( \overline{\bm{Y}_o(s)} \mid \tT_o = u, \bV_o )$ with $u \leqslant \tau_{\text{max}}$, and another for $P( \overline{\bm{Y}_o(s)} \mid \tT_o > \tau_{\text{max}}, \bV_o )$, the latter corresponding to long-term survivors (LTS; \citealt{SKWang2020Biostat}), i.e., those with $\tT_o > \tau_{max}$.

In practice, we can use the estimated marginal distribution of $\tT$ to guide whether a separate longitudinal sub-model for the LTS is needed. If the Kaplan–Meier curve for $\tT$ drops near zero, the LTS group is likely negligible, and a single sub-model suffices; the extrapolation error is expected to be small. We refer to it as the \emph{BJM with extrapolation (BJM-EX)}, where $P(\tT = u \mid \bV)$ and $P( \overline{\bm{Y}_o(s)} \mid \tT_o = u, \bV_o )$ as functions of $s$ are extrapolated beyond the observed data range. If the LTS group is non-negligible, a second longitudinal sub-model is required, resulting in the \emph{two-part BJM (BJM-TP)}. For BJM-TP, predictions are limited to times $s < \tau_{\text{max}} - \Delta$, since no terminal events are observed beyond $\tau_{\text{max}}$, and no extrapolation is performed. In contrast, BJM-EX allows prediction beyond $\tau_{\text{max}}$ by relying on the extrapolation.

\textbf{Remark 2.1 (Stacked building blocks for the joint distribution).}  
Traditional regression model building begins with specifying a data-generating mechanism, formulating a compatible model, and deriving the likelihood for estimation and inference. Our approach takes the reverse route: we start with the likelihood and decompose it into components, each amenable to standard statistical methods available in typical software. This modular construction enables the joint model to be built by ``stacking'' basic building blocks, allowing for stable and efficient computation, scalability to multivariate longitudinal data, and direct implementation by existing software (Section~\ref{sec:training}).   

% Not all likelihood factorizations result in such ease of implementation. For instance, the shared random effects joint model expresses the joint distribution as
% While each distribution under the integral sign is modeled by a standard statistical method, they cannot be estimated as separate building blocks because they share the random effects. The joint likelihood involves integration over latent variables, often leading to substantial computational challenges.

\textbf{Remark 2.2 (Reversing the conditional relationship).} The shared random effects model (SJM) was originally developed to explain the association between longitudinal and survival data or to adjust for informative dropout \citep{Albert2018}. It uses the likelihood factorization
\begin{equation}
P(\bY_o, \tT_o \mid \bV_o) = \int P(\tT_o \mid \bV_o, \bb) \, P(\bY_o \mid \bV_o, \bb) \, P(\bb \mid \bV_o) \, d\bb ,    
\label{eq:SJM.likelihood}
\end{equation}
where $\bb$ denotes shared random effects. The SJM was later adapted for the dynamic prediction problem in (\ref{DP_risk}) \citep{Taylor2005, Yu2008, Rizopoulos2011}. While real-world data generation is more complex than any model can fully capture, parsimonious models aligned with plausible causal mechanisms (e.g., modeling outcomes conditional on past covariates) offer interpretability and have contributed to the widespread adoption of the SJM. In contrast, the BJM was designed primarily as a prediction tool; hence, it does not reflect the data-generating mechanism. In predictive modeling, reversing the temporal order of the outcome and predictor in conditional distributions is common, since association --- rather than causal implication ---is all that's needed for accurate prediction. For example, the Naive Bayes classifier models the covariate distribution given the outcome, then applies Bayes rule to predict the outcome. The BJM adopts a similar principle. Such reverse modeling is also prevalent in various other statistical applications (Section \ref{subsec:literature}). A recent comprehensive discussion of modeling longitudinal data with reversed time scale can be found in \citet{dempsey2018survival}, who called such models a \textit{revival model}. The parameters in the BJM can be interpreted as quantifying the association between survival and longitudinal trend (\textbf{Remark 3.1} in Section \ref{sec:BJM.comprisk}). Despite their apparent differences, the SJM and BJM are equivalent in certain situations and are expected to produce the same prediction (Section \ref{subsec:connection}).

\subsection{\textbf{Sub-models for the components of the factorized likelihood}}
\label{subsec:survival.longitudinal.submodels}

The likelihood component $P(\tT_o \mid \bV_o)$ can be modeled using any appropriate survival regression with time-invariant covariates, such as the Cox model, the Weibull model, the accelerated failure time model, etc. Under the independent censoring assumption, the model can be estimated from the training data $\{ T_i, \bV_i; i=1,2,...,n \}$ using standard survival analysis software, and the longitudinal data are not involved in this step. Model diagnosis procedures for these models are well-developed to mitigate model misspecification. The survival sub-model for $P(\tT_o \mid \bV_o)$ also produce an estimate of $P(\tT_o > \tau_{max} \mid \bV_o)$ for use in the BJM-TP. 

The likelihood component $P(\bY_o \mid \tT_o = u, \bV_o)$ for $u \in (0, \infty)$ (BJM-EX) or $u \in (0, \tau_{max})$ (BJM-TP) can be modeled using any appropriate multivariate linear mixed model with the time-invariant covariates $\tT_o$ and $\bV_o$ (the measurement times, $\bt_i$, are also included). Throughout this paper, we focus on continuous biomarkers in $\bY_o$. The extension to a mixture of continuous and categorical biomarker variables is discussed in Section \ref{sec:Discussion}.      

The likelihood component $P(\bY_o \mid \tT_o > \tau_{max}, \bV_o)$ in BJM-TP needs to be estimated with a second longitudinal sub-model, defined on the LTS subjects ($\tT_i > \tau_{max}$). The multivariate linear mixed model is a natural choice. The covariates in this model do not include $\tT$ because this model describes the overall distribution of the longitudinal data from LTS subjects, which comprise those with various $\tT$ values. In other words, the model conditions on $\tT_o > \tau_{max}$, but any specific $\tT_o$; this is different from the longitudinal sub-model for the non-LTS. Specifically, the second term in (\ref{eq:integral.split}) equals to
\begin{equation}
    P( \overline{\bm{Y}_o(s)} , \tT_o > \tau_{\text{max}} \mid \bV_o ) = \int_{\tau_{max}}^{\infty} P( \overline{\bm{Y}_o(s)} \mid \tT_o = u, \bV_o ) P( \tT_o = u \mid \bV_o ) du . 
    \label{eq:second.term.TP}
\end{equation}
It is determined by two functions, $P( \overline{\bm{Y}_o(s)} \mid \tT_o = u, \bV_o )$ and $P( \tT_o = u \mid \bV_o )$, for $u \in (\tau_{max}, \infty)$. Unless an extrapolation is used, their values on $u \in (0, \tau_{max})$, which are identifiable and part of the first term in (\ref{eq:integral.split}), do not inform (\ref{eq:second.term.TP}). Therefore, a separate sub-model for the LTS with its own parameters is needed in BJM-TP.

In the exhibition above, we discussed model formulation for BJM and also the estimation of the survival sub-model. The estimation of longitudinal sub-models requires accounting for censoring. The determination of whether a subject belongs to the LTS also requires consideration for censoring because subjects censored before $\tau_{\max}$ may still belong to the LTS. Section~\ref{sec:BJM.comprisk} discusses the necessary censoring adjustment in model estimation.

\subsection{\textbf{Related literature}}
\label{subsec:literature}

The basic ideas of the BJM were introduced by \citet{Shen2021}, who focused on the estimand (\ref{DP_risk}) and a two-part BJM formulation. This paper extends that work with a more comprehensive discussion and several new developments. Chapter 8.1 of \citet{vanHouwelingen2011} proposed that any joint distribution of $\tT$ and $\bY$ could be used for prognostic prediction. They also introduced the term ``backward modeling,'' though the discussion was limited to using uncensored data. We further speculate that the selection model (e.g., Chapter 18 of \citealt{Handbook.Longitudinal.Data}), another joint modeling approach for missing data, may be applicable in this context. However, the SJM and its variants are well-established and widely used, and for an alternative joint model to be useful, it must offer advantages the SJM lacks. In our view, the BJM stands out due to its computational simplicity and ease of implementation. In addition, \citet{putter2022landmarking} proposed some similar ideas as the BJM to build a revival model, such as reversing the time scale (\textbf{Remark 2.2}) and using the long-term survivors. However, their revival model is used to estimate a predictable time-dependent covariate process, which is then used in a landmark modeling-based dynamic prediction method.   

The ``backward'' factorization has also been employed in other areas beyond prediction, such as pattern mixture models for missing data \citep{Little1993, Hogan2004}, conditional linear models for informative dropout \citep{Wu.Bailey.1989}, longitudinal quality-of-life modeling near terminal events \citep{LiZigang2012, LiZigang2017, Wu2023}, and describing individual trajectory patterns stratified by death \citep{Kurland2009, Li2018JASA, Kong2017JASA, Wang2023JASA, SKWang2020Biostat, ShikunAOAS2023, SingleIndex2023}. In all these cases, reversing the conditional relationship (\textbf{Remark 2.2}) led to interpretable model parameters. We illustrate the interpretation of the BJM in Sections \ref{subsec:submodel.longitudinal.comprisk}.

\subsection{\textbf{Connection with other joint models}}
\label{subsec:connection}

While the SJM and BJM are different models, they have overlap, i.e., some SJMs can be written as BJMs. In such situations, the SJM and BJM are equivalent and expected to produce the same prognosis. The likelihood of an SJM in (\ref{eq:SJM.likelihood}) always implies that a backward conditional distribution exists:
\begin{equation}
P(\bY_o \mid \tT_o , \bV_o) = \dfrac{ \int P(\tT_o \mid \bV_o, \bb) \, P(\bY_o \mid \bV_o, \bb) \, P(\bb \mid \bV_o) \, d\bb }{ \int P(\tT_o \mid \bV_o, \bb) \, P(\bb \mid \bV_o) \, d\bb } .  
\end{equation}
However, we refer to it in this paper as a BJM only when this conditional distribution takes the form of a multivariate linear mixed model, because this is the formulation discussed here. Of note, this formulation still allows the mean of $\bY_o$ to be linearly associated with nonlinear transformations of $\tT_o$ and $\bV_o$. Consider the following SJM:
\begin{equation}
\begin{aligned}
    \bY_i = & ~ \bX_i \bbeta + \bZ_i \bb_i + \bepsilon_i ~, \\
    H( \tT_i ) = & ~ \balpha^T \bV_i + \bgamma^T \bB_i \bb_i + \xi_i ~. 
    \label{eq:overlap}
\end{aligned}
\end{equation}
The first equation defines the longitudinal sub-model, a multivariate linear mixed model. Here, $\bX_i$ and $\bZ_i$ are design matrices for the fixed and random effects, respectively, both known functions of $\bV_i$ and $\bt_i$. The random effects $\bb_i \sim N(\bs 0, \bSigma_{\bb})$, and the residual errors $\bepsilon_i \sim N(\bs 0, \bSigma_{\bepsilon})$, are mutually independent. The second equation specifies the survival sub-model, with $H(\cdot)$ denoting a monotone increasing transformation, $\bB_i$ is a known design matrix, and $\xi_i \sim N(0, \sigma^2_{\xi})$ is an independent residual term. The random vectors $\bb_i$, $\bepsilon_i$, and $\xi_i$ are assumed to be independent. Several well-known joint models are special cases of (\ref{eq:overlap}). For instance, \citet{LiuMengLing2007} assumes an unknown $H(\cdot)$ with $\sigma^2_{\xi} = 1$; \citet{Schluchter1992} uses a logarithmic $H(\cdot)$ with an unknown $\sigma^2_{\xi}$. Under (\ref{eq:overlap}), the joint distribution $(\bY_i^T, H(\tT_i)^T \mid \bV_i)$ is multivariate normal, implying that $P(\bY_i, \tT_i \mid \bV_i)$ admits a BJM representation: $\bY_i \mid \tT_i, \bV_i$ follows a multivariate linear mixed model, and $\tT_i \mid \bV_i$ follows a linear transformation model for survival \citep{Cheng1995LTM}.

In an SJM where the hazard of the terminal event is proportional to $\bV_i$ and $\bb_i$ (e.g., \citealt{Liu2008three.models}), $\tT_i \mid \bV_i, \bb_i$ follows a Cox proportional hazards model with time-independent covariates, implying that $\xi_i$ follows a standard extreme value distribution with CDF $F(s) = 1 - \exp\{-\exp(s)\}$. Suppose this distribution is approximated by a finite mixture of $K$ known normal distributions with mixing probabilities $(p_1, \ldots, p_K)$: specifically, $\xi_i = \sum_{k=1}^K \xi_{ki} \bI\{ c_i = k \}$, where $\xi_{ki} \sim N(\mu_k, \sigma_k^2)$ and $c_i \sim \mathrm{Multinomial}(p_1, \ldots, p_K)$ are independent. Then, $\tT_i \mid \bV_i$ still follows a linear transformation model, and $\bY_i \mid \tT_i, \bV_i, c_i = k$ each follows a multivariate linear mixed model. Marginalizing over $c_i$, $\bY_i \mid \tT_i, \bV_i$ becomes a multivariate linear mixed model with non-Gaussian random effects and residuals. Despite this added complexity, the joint distribution $P(\bY_i, \tT_i \mid \bV_i)$ retains a BJM structure. 

When the hazard of the terminal event is non-proportional to $\bV_i$ and $\bb_i$, the SJM may not be representable as a BJM. For instance, the survival sub-model in an SJM is a Cox model with hazard $\lambda_0(t)\exp\{ \bV_i^T \bgamma_1 + \gamma_2 m_i(t) \}$ (e.g., \citealt{Andrinopoulou2021}), where $m_i(t)$ denotes subject $i$'s true biomarker trajectory at time $t$ and the observed data follow $Y_i(t) = m_i(t) + \epsilon_i(t)$. However, \citet{AlbertShih2010} argued that the conditional distribution $\bb_i \mid \bV_i, \tT_i$ can be approximated by a normal distribution so that $\bY_i \mid \bV_i, \tT_i$ is modeled by a linear mixed model. In the framework of this section, the SJM has an approximate BJM representation. That argument was made when the longitudinal sub-model is a random intercept and slope model: $m_i(t) = b_{0i} + b_{1i} t$, and the measurement times $\bt_i$ are on a discrete scale.   

The SJM does not require a sub-model for the LTS, whereas the BJM sometimes does. This is because the SJM specifies a hazard model for $P(\tT_o \mid \bV_o, \bb)$ in (\ref{eq:SJM.likelihood}), leveraging the deterministic link between hazard and survival functions to simplify likelihood computation under censoring. However, the inclusion of random effects $\bb$ in the hazard leads to intractable numerical integration. In contrast, the BJM models the joint distribution of $\bY_o$, $\tT_o$, and $\bV_o$ directly, avoiding the hazard function. Its random effects capture individual biomarker trajectories and can often be integrated out analytically --- though this may require an additional sub-model for the LTS to be valid.

\subsection{\textbf{Non-informative observational times and independent censoring}}
\label{subsec:censoring}
The proposed BJM shares two standard assumptions with most existing joint models: non-informative observation times and independent censoring (Section~\ref{subsec:overview.DP}). To our knowledge, no dynamic prediction methods currently utilize informative observation times $\bt_i$ to improve prediction, though in principle, this could be addressed through a joint model of $\bY_i$, $\bt_i$, and $\tT_i$ given $\bV_i$. Compared to standard joint models of longitudinal ($\bY_i$) and survival ($\tT_i$) data, this would require modeling longitudinal, survival, and recurrent event data ($\bt_i$) jointly \citep{Liu2008three.models}.

Most published SJMs assume independent censoring \citep{WulfsohnTsiatis1997, tsiatis2004joint, Handbook.Longitudinal.Chap15}. For example, in longitudinal cohort studies with informative censoring --- the original motivation for SJMs \citep{Albert2018} --- the informative censoring event plays the role of $\tT_i$ in the BJM, while the non-informative censoring event corresponds to $C_i$ \citep{Schluchter1992, LiuMengLing2007}. \citet{Handbook.Longitudinal.Chap15} suggested extending the SJM framework by introducing a third sub-model for the dependent-censoring process, leading to a three-component structure: $\bY_i \mid \bV_i, \bb_i$ (repeated-measurement sub-model), $\tT_i \mid \bV_i, \bb_i$ (hazard-of-death sub-model), and $C_i \mid \bV_i, \bb_i$ (hazard-of-censoring sub-model), and these components are conditionally independent given the shared random effects $\bb_i$. Marginalizing over $\bb_i$ induces dependence among $\bY_i$, $\tT_i$, and $C_i$ given $\bV_i$. With a similar idea but different context, \citet{Wu2023} studied a joint model that accommodates correlation among death, an informative censoring event, and a longitudinal variable by using a shared random effect in three sub-models. Despite its conceptual appeal, the hazard-of-censoring sub-model has seen limited adoption in both methodological and applied SJM literature. Notably, if $\tT_i$ and $C_i$ are dependent, the primary analysis of the longitudinal cohort study (e.g., comparing treatment effects on survival) becomes statistically unidentifiable using standard survival methods. Consequently, joint modeling, typically reserved for secondary analysis, cannot substitute for a valid primary analysis when independent censoring is violated. 

The BJM can, in principle, be extended to relax the assumptions on non-informative observational times and independent censoring. For example, \citet{Wu2023} employed a reverse time scale (\textbf{Remark 2.2}), and this approach may be modified for use in the BJM problems. However, such developments are beyond the scope of this paper and are left for future development of the BJM.

\section{The Extension to Competing Risk}
\label{sec:BJM.comprisk}

We now address the fourth statistical challenge in Section \ref{subsec:motivating.applications}, the BJM for competing risk outcomes (crBJM). The notation, model formulation and estimation procedure for the competing risk context are similar to the non-competing risk case and include the latter as a special case. Since Sections \ref{sec:intro} and \ref{sec:BJM.basic} already covered much content for the non-competing risk case, this section will focus only on the differences.  

Suppose that the terminal event can be one of $J$ event types, denoted by $\tD_i \in \{1, 2, ..., J\}$. $J=1$ corresponds to the non-competing risk setting. In our motivating data application, $J=2$, and the terminal event is either ESRD ($j=1$) or death ($j=2$). The notations for the true and observed terminal event times, censoring time and indicator are the same. The observed event type is $D_i = \delta_i \tD_i \in \{0, 1, ..., J\}$, where $\delta_i$ is the censoring indicator defined in Section \ref{subsec:overview.DP}. 

The dynamic prediction formula (\ref{DP_risk}) can be modified for predicting the probability of having the terminal event of type $j$ in the next $\Delta$ years (or other time unit), given the subject is at-risk by time $s$ and the observed history, 
\begin{equation}
\begin{aligned}
	 & ~ P(s < \tT_o \leqslant s + \Delta, \tD_o = j | \overline{\bY_o(s)}, T_o > s, \bV_o) \\
    = & ~ \dfrac{P( \overline{\bm{Y}_o(s)},  s<\tilde{T}_o \leqslant s +\Delta,\tD_o = j |\bV_o) }{P( \overline{\bm{Y}_o(s)},\tT_o > s| \bV_o)} . 
    \label{DP_risk.cr}
\end{aligned}
\end{equation} 
The dynamic prediction for future longitudinal data remains the same as (\ref{DP_eGFR}) because the longitudinal data exist as long as the terminal event does not occur, regardless of the event type. 

Analogous to (\ref{eq:factorize}), the likelihood factorization of the crBJM is
\begin{equation}
P(\bY_o, \tT_o, \tD_o \mid \bV_o) = P(\tT_o, \tD_o \mid \bV_o) P(\bY_o \mid \tT_o, \tD_o, \bV_o)~.~ 
\label{eq:factorize.cr}
\end{equation}
The crBJM comprises of a sub-model for $P(\tT_o, \tD_o \mid \bV_o)$ and sub-models for $P(\bY_o \mid \tT_o, \tD_o, \bV_o)$. Since the denominators of both estimands (\ref{DP_risk.cr}) and (\ref{DP_eGFR}) involve only $ P( \overline{\bm{Y}_o(s)},\tT_o > t| \bV_o) $ ($t \geqslant s$), and not the event type ($\tD_o$), the same derivation as in Section \ref{subsec:twoBJM} applies and justifies the use of an extrapolation-based longitudinal sub-model (crBJM-EX) or a two-part sub-model (crBJM-TP). The model formulation is similar between crBJM-EX and the longitudinal sub-model for the non-LTS subjects in crBJM-TP. For the LTS, a separate longitudinal sub-model is needed and that model does not involve the event type and event time. See the discussion surrounding equation (\ref{eq:second.term.TP}).

\subsection{\textbf{The sub-model for the competing risk outcomes}}
\label{subsec:submodel.comprisk}

Any competing risk regression model with time-invariant covariates can be used to fit $\{ T_i, D_i, \bV_i; i=1,2,...,n \}$ and estimate $P( \tT_o,\tD_o | \bV_o )$ in (\ref{eq:factorize.cr}). Common choices include the cause-specific hazard models, the Fine-Gray sub-distribution hazard model, or the vertical modeling \citep{Nicolaie2010}, among others. We used vertical modeling in the numerical studies because it directly models $P( \tT_i | \bV_i )$, which is convenient for predicting longitudinal data in (\ref{DP_eGFR}). We compared the crBJMs with vertical modeling and with cause-specific hazard models, and the predictive performances were similar in our data application.

The likelihood of a vertical model for $P(\tT_i, \tD_i | \bV_i)$ is factorized as the likelihood product of $P( \tT_i | \bV_i )$ and $P(\tD_i | \tT_i, \bV_i )$. The former can be formulated as a semi-parametric Cox model, a parametric Weibull model, or any other survival regression model that fits the data well. The latter is a multinomial regression model according to \citet{Nicolaie2010}. Specifically, $P(\tD_i = j | \tT_i = t, \bV_i) = \frac{\exp( \bB(t)^T \bxi^{(\tT)}_{j} + \bV_i^T \bxi^{(\bV)}_{j} ) }{ \sum_{j = 1}^{J} \exp( \bB(t)^T \bxi^{(\tT)}_{j} + \bV_i^T \bxi^{(\bV)}_{j} ) }, j = 1,..., J$. The $\bB(t)$ is a vector of pre-specified regression spline basis to accommodate the nonlinear effect of $\tT_i$ on the event types. When $J=2$, this is a logistic regression with partially linear specification for the covariates $\tT_i$ and $\bV_i$. The likelihood-based estimation of vertical modeling follows \citet{Nicolaie2010} and is omitted here.

\subsection{\textbf{The sub-model for longitudinal data given competing risk outcomes}}
\label{subsec:submodel.longitudinal.comprisk}

For the crBJM-EX, $P( \bY_o | \tT_o, \tD_o, \bV_o )$ is estimated by a multivariate linear mixed model, with $T_i$, $D_i$, $\bV_i$, $\bt_i$ and their interactions as fixed-effect covariates and with random effects (e.g., intercept and slope) to account for the longitudinal correlation and the correlation among different biomarkers. The sub-model of the $m$-th longitudinal biomarker ($m=1, 2,..., M$) can be formulated in the following general form:  
\begin{equation}\label{BJM_model}
	Y_{mi}(t) = \sum_{l = 0}^L \vartheta_{ml}(\bV_{i}, \phi(\tT_i), \tD_i ) g_l(t) + \bu_{mi}(t)^T \bb_{mi} + \epsilon_{mi}(t) ~. 
\end{equation}
$Y_{mi}(t)$ is subject $i$'s $m$-th biomarker, measured at time $t$. In the training data, $t \in \bt_{mi}$. The first term describes the conditional mean trajectory (CMT) given the time-invariant variables $\bV$, $\tT$ and $\tD$. The pre-specified transformation $\phi(.)$, e.g. $\log$, may be used to reduce the skewness of the time variable $\tT_i$. The conditioning on $\tT_i$ renders the CMT an interpretation as the mean trajectory among subjects with the same trajectory lengths. It does not lead to proper interpretation to average the longitudinal biomarker trajectories of different lengths when the follow-up duration is correlated with longitudinal data, a defining feature of joint modeling problems that is sometimes called informative dropout. The $g_l(t), l = 0, 1, ..., L$ are basis functions that characterize the longitudinal shape of the CMT, with $\vartheta_{ml}(\bV_{i}, \phi(\tT_i), \tD_i )$ being the regression coefficient of the basis, expressed in a varying-coefficient model form as a function of $\bV$, $\tT$ and $\tD$. 

The second term includes random effects that describe the subject-specific, mean-zero deviation from the CMT. The random intercept ($\bu_{mi}(t) = 1$), random intercept plus slope ($\bu_{mi}(t) = (1, t)^T$), or other nonlinear random effect terms can be used. The random effects from all $M$ biomarkers $\bm{b}_{i} =(\bm{b}_{1i}^T, \bm{b}_{2i}^T,..., \bm{b}_{Mi}^T )^T$ follows a multivariate normal distribution with mean zero and covariance matrix $\bm{\Omega}$. When model (\ref{BJM_model}) has random intercept and slope, $\bm{\Omega}$ is a $2M \times 2M$ matrix that is not restricted to be block diagonal, allowing the biomarkers to be correlated. We can impose further assumptions on this correlation matrix to reduce the number of unknown parameters in the model. The residuals $\varepsilon_{mi}(t)$ are independent noises with mean zero and variance $\sigma_{e,m}^2$. Extensions to residuals with auto-correlation are straightforward. 

The crBJM-TP has an extra longitudinal sub-model for the LTS: 
\begin{equation}\label{LTSmodel}
	Y_{mi}(t) = \sum_{l = 0}^L \vartheta_{ml}^e(\bV_{i} ) g_l(t) + \bu_{mi}^e(t)^T \bb^e_{mi} + \epsilon_{mi}^e(t) ~,~ \tT_i > \tau_{max}. 
\end{equation}
The superscript $e$ symbolizes quantities in the ``extra'' LTS model. The model formulation and interpretation are similar to (\ref{BJM_model}) but with a few differences. First, model (\ref{LTSmodel}) does not involve $\tT_i$ and $\tD_i$ in the covariates because the LTS includes all subjects who survive beyond $\tau_{max}$ regardless of their actual $\tT_i$ and $\tD_i$. Second, since the LTS is a heterogeneous group, models (\ref{LTSmodel}) and (\ref{BJM_model}) cannot have the same random effect variance or residual variance. Presumably, the variances are larger in model (\ref{LTSmodel}). The sub-model for the non-LTS does not inform the sub-model for LTS, which justifies the latter to have independent parameters of its own. See equation (\ref{eq:second.term.TP}) and discussions therein. 

\textbf{Remark 3.1 (Interpretation).} The parameters in the longitudinal sub-model of a BJM characterize how the shape of longitudinal trajectories varies across subjects in different survival strata. For example, consider model (\ref{BJM_model}) with a linear CMT: $L=1$, $g_0(t)=1$, and $g_1(t) = t$. The coefficients $\vartheta_{ml}(\bV_i, g_1(\tT_i), \tD_i)$ for $l=0,1$ represent the mean intercept and slope for subjects sharing the same $\bV_i$, $\tT_i$, and $\tD_i$. Suppose they are modeled as:
\begin{align*}
\vartheta_{m0}(\bV_i, \phi(\tT_i), \tD_i) &= \bbeta_{m1}^T \bV_i + \nu_{m0} \tD_i + \nu_{m2} \phi(\tT_i) + \nu_{m4} \phi(\tT_i) \tD_i, \\
\vartheta_{m1}(\bV_i, \phi(\tT_i), \tD_i) &= \bbeta_{m2}^T \bV_i + \nu_{m1} \tD_i + \nu_{m3} \phi(\tT_i) + \nu_{m5} \phi(\tT_i) \tD_i.
\end{align*}
These expressions describe how the average trajectory changes with covariates, event time, and event type. For instance, $\nu_{m0}$ captures a vertical shift in mean trajectories between event types for subjects with a fixed $\tT$, while a non-zero $\nu_{m4}$ indicates that the amount of shift depends on $\tT$. Similar interpretations of BJM parameters have been used in other applications, such as modeling medical costs or terminal decline in quality of life (see Section~\ref{subsec:literature}).

\textbf{Remark 3.2 (Extrapolation error).} When both the longitudinal sub-model and survival sub-model are parametric, there is no extra extrapolation step in BJM-EX. When the survival sub-model is semi-parametric, such as a Cox model, the tail distribution of $\tT \mid \bV$ is unidentified. In such a situation, any reasonable extension of the estimated part of the survival distribution can be used. We have used a linear extension of the log baseline hazard, which is equivalent to appending an exponential distribution at the tail. 

\textbf{Remark 3.3 (Semi-competing risk vs. competing risk).} This paper focuses on competing risks, characterized by the time to the composite event ($\tT$) and the event type ($\tD$). Semi-competing risks, where one event precludes observation of the other but not vice versa, require different methods and are beyond the scope of this paper \citep{Peng2007}. In our motivating application, the outcomes are death and ESRD. Although death can occur after ESRD, ESRD is typically treated as a terminal event in CKD studies, as disease management and patient health status have fundamental changes after its onset (Section \ref{subsec:motivating.applications}).

\section{Model Training and Prediction Accuracy}
\label{sec:training}

The crBJM can be estimated from the training data in two steps. In the first step, we fit the competing risk regression model (Section \ref{subsec:submodel.comprisk}). Many standard software packages are available. In the second step, we estimate the longitudinal sub-models in Section \ref{subsec:submodel.longitudinal.comprisk}, and this is the focus of this section. 

We propose two approaches. Section \ref{subsec:CCA} presents a simple method that can be conveniently implemented with standard linear mixed model software. This method only uses training data subjects whose terminal event times are observed and those still at-risk by $\tau_{max}$. Since the censored data are not used, this approach is referred to as complete-case analysis (CCA). A remarkably useful feature of the crBJM is that the CCA produces consistent estimation, albeit less efficient. The CCA of crBJM enables quick and convenient exploratory analysis, model building, model assumption diagnosis, and preliminary assessment of prediction accuracy. Once the model specification is finalized, we can use the fully likelihood-based analysis in Section \ref{subsec:EM.full.data} to obtain efficient estimators as the final results. Using the CCA estimator as the initial values is helpful for the iterative likelihood optimization. Since these initial values are already consistent, the iterations often converge quickly.

\subsection{\textbf{Complete-case analysis (CCA)}}
\label{subsec:CCA}

To see why the parameters of longitudinal sub-model can be consistently estimated with the complete cases in the training data (i.e., $D_i \neq 0$), we observe that, for any $j=1,2,...,J$, $P( \bY | \tT = t, \tD = j, \bV ) = P( \bY | \tT = t, C > t, \tD = j, \bV ) = P( \bY | T = t, D = j, \bV )$. The simple derivation is justified by the independent censoring (Section \ref{subsec:censoring}). It holds at any $t \in (0, \infty)$ for the crBJM-EX, and any $t \in (0, \tau_{max})$ for the crBJM-TP. $P( \bY | T = t, D = j, \bV )$ can be estimated via a multivariate linear mixed model according to the specification in (\ref{BJM_model}). Additionally, since $P( \bY | \tT > \tau_{max}, \bV ) = P( \bY | \tT > \tau_{max}, C > \tau_{max}, \bV ) = P( \bY | T > \tau_{max}, \bV )$ in the crBJM-TP, we can estimate the multivariate linear mixed model (\ref{LTSmodel}) from the subset of training data with $T_i > \tau_{max}$. Consistent estimation of CCA has been demonstrated in other published models with a similar likelihood factorization \citep{Li2018JASA, SKWang2020Biostat, Shen2021, Wang2023JASA}. 

% \textbf{Remark 4.1 (Model checking).} The simple, quick and consistent estimation by the CCA suggests

\subsection{\textbf{EM algorithm}}
\label{subsec:EM.full.data}

This section describes the EM algorithm for the longitudinal sub-model estimation. Since it uses both censored and non-censored data, the estimation is more efficient than the CCA \citep{Li2018JASA, SKWang2020Biostat, Shen2021}. The censored $\tT_i$ and $\tD_i$ are treated as missing data in the complete data log-likelihood. Since the competing risk sub-model are estimated consistently in a prior step, the method is pseudo maximum likelihood estimation \citep{pseudoMLE}.

Let $\bTheta$ denote the parameters of the longitudinal sub-model. With crBJM-TP, $\bTheta$ includes additionally the parameters $\bTheta^e$ from the LTS sub-model (\ref{LTSmodel}). The estimated parameters from the survival sub-model are denoted by $\hat{\bTheta}_{surv}$. Since $\hat{\bTheta}_{surv}$ is fixed throughout the EM iterations, we omit it for the formula below. In the following, we mainly describe the EM algorithm of the crBJM-EX, but will point out the changes needed when the crBJM-TP is used. The complete data log-likelihood is $  \sum_{i = 1}^{n} \left\{ \log P(\bY_i | \tT_i, \tD_i, \bV_i; \bTheta) +  \log P(\tT_i, \tD_i | \bV_i) \right\} $.  The observed data of all subjects in the training dataset is collected in $\{\bY, \bm{T}, \bm{\delta}, \bD, \bV \}$. The EM algorithm uses the consistent estimators from the CCA as the initial values in the first EM iteration. At the beginning of the $q$-th iteration ($q=1,2, ...$), the estimated parameter is denoted by $\hat{\bm{\Theta}}^{(q)}$. 

In the E-step of the $q$-th iteration, we calculate the conditional expectation of the complete data log-likelihood as in (\ref{Expectation_CDL}). This expectation is taken over the conditional distribution of the missing data (censored $\tT_i$ and $\tD_i$) given the observed data, with parameters in this distribution set at $\hat{\bm{\Theta}}^{(q)}$. 
\begin{equation}
	\begin{aligned}
		& \sum_{i = 1}^{n} E_{ ( \tT_i, \tD_i ) }\left\{ \log P(\bY_i | \tT_i, \tD_i, \bV_i; \bTheta^{(q + 1)}) +  \log P(\tT_i, \tD_i | \bV_i) \Big| \bY, \bm{T}, \bm{\delta}, \bD; \hat{\bTheta}^{(q)} \right\}  \\
		\propto & \sum_{i = 1}^{n} \sum_{j=1}^J \int_{T_i}^{\infty} \left\{ \log P(\bY_i | \tT_i, \tD_i = j, \bV_i; \bTheta^{(q + 1)}) \right\} P( \tT_i, \tD_i = j | \bY, \bm{T}, \bm{\delta}, \bD; \hat{\bTheta}^{(q)} ) d\tT_i 
		\label{Expectation_CDL}
	\end{aligned} 
\end{equation}
There is always a one-dimensional integral in (\ref{Expectation_CDL}), regardless of how many longitudinal biomarker variables are involved. This integral is needed only for censored subjects. The integration in (\ref{Expectation_CDL}) goes from $T_i$ to $\infty$. With crBJM-TP, this calculation is split into two parts, analogously to (\ref{eq:integral.split}). The first integrates from $T_i$ to $\tau_{max}$ using model (\ref{BJM_model}) and the second uses the LTS model (\ref{LTSmodel}).  

The integral in (\ref{Expectation_CDL}) can be calculated with the trapezoidal rule. If $T_i$ is a censoring time, we assume that $\tT_i$ falls in one of a series of pre-specified connected intervals, denoted by $\{ \mathbb{L}_{ik} = [a_{ik}, b_{ik}) ; k =1,...,K_i \}$. The intervals are connected (e.g., $b_{ik} = a_{i(k+1)}$) and starts from $T_i$ ($a_{i1}=T_i$). With crBJM-EX, we can set the ending time of the last interval at a very large survival time (e.g., 100 year old). With crBJM-TP, the last interval is $[ \tau_{max}, \infty)$ and the LTS model is used on this interval. A moderate $K_i$ can produce accurate results when the residual survival beyond $T_i$ is not very large or when the integrand is not excessively wiggly \citep{Li2018JASA}. The target function (\ref{Expectation_CDL}) of crBJM-EX then becomes:
\begin{equation}
	\begin{aligned}
		\sum_{i = 1}^{n} \sum_{k=1}^{K_i} \sum_{j=1}^J \left\{ \log P(\bY_i | \tT_i = \dfrac{a_{ik} + b_{ik}}{2}, \tD_i = j, \bV_i; \bTheta^{(q + 1)}) \right\} \times \\
        P( \tT_i \in \mathbb{L}_{ik}, \tD_i = j | \bY, \bm{T}, \bm{\delta}, \bD; \hat{\bTheta}^{(q)} ). 
		\label{target.Estep}
	\end{aligned} 
\end{equation}
With crBJM-TP, the target function is the same except that the terms corresponding to the last interval ($\mathbb{L}_{iK_i}$) are replaced by 
$ \left\{ \log P(\bY_i | \tT_i \in \mathbb{L}_{iK_i}, \bV_i; \bTheta^{(q + 1)}) \right\} \linebreak P( \tT_i \in \mathbb{L}_{iK_i} | \bY, \bm{T}, \bm{\delta}, \bD; \hat{\bTheta}^{(q)} ) $. 

In the M-step, (\ref{target.Estep}) is maximized with respect to $\bTheta^{(q + 1)}$, producing an updated estimator $\hat{\bTheta}^{(q + 1)}$ that will enter the next ($(q+1)$-th) iteration. The only terms containing the unknown parameters are $\log P(\bY_i | \tT_i = (a_{ik} + b_{ik})/2, \tD_i = j, \bV_i; \bTheta^{(q + 1)}) $. These are log of multivariate normal density functions, which are quadratic forms of the regression coefficients in the longitudinal sub-models.  $P( \tT_i \in \mathbb{L}_{ik}, \tD_i = j | \bY, \bm{T}, \bm{\delta}, \bD; \hat{\bTheta}^{(q)} )$ is calculated in the E-step and treated as fixed in the maximization of the M-step. Therefore, the regression coefficients have closed-form solutions (the variance-covariance matrices of the random effects do not; see below). This feature, in conjunction with using the consistent estimators from the CCA as the initial value of the EM algorithm, speeds up the computation considerably and improves the robustness of the iterations. 
% We have observed a 100\% convergence rate with the EM and the computing time is much shorter compared to the shared random effects joint model. 

\textbf{Remark 4.1 (Number of parameters with multivariate longitudinal data).} The number of regression coefficients in the crBJM increases linearly with $M$, the number of longitudinal biomarkers. However, the number of parameters in the variance-covariance matrix of the random effects goes up quadratically with $M$. This poses a challenge not just for BJM, but also for any joint models that aim to estimate the multivariate distribution of the longitudinal data \citep{Hashemi2025}. The consistent estimation by the CCA suggests a solution to avoid this computational burden: during the EM algorithm, we fix the variance-covariance matrix of the random effects and residuals at their CCA estimators. This solution is justified by the pseudo maximum likelihood estimation \citep{pseudoMLE}. Another solution to this problem is to use a pairwise approach \citep{Fieuws2006, AlbertShih2010}, i.e., fitting the BJM with two longitudinal biomarkers at a time, and estimating the multiple pairwise longitudinal sub-models in parallel computing. These two solutions are feasible due to the specific modeling structure of the BJM and are not easily implementable in other joint models.

\subsection{\textbf{Variance estimation}} 
\label{subsec:variance.estimation}

We use the bootstrap to estimate variances. Specifically, we repeatedly sample subjects with replacement from the training dataset to create bootstrap samples. Each sample is analyzed using the crBJM procedure to obtain point estimates of the model parameters. This process is repeated 200 times, and the variance of each parameter is estimated as the sample variance of its 200 bootstrap estimates.

For a new subject in the validation dataset, predictions from (\ref{DP_risk}) or (\ref{DP_eGFR}) depend on the estimated crBJM parameters and the subject's data prior to the prediction time. The variance of the prediction can be approximated using the Delta method to account for the variability in the estimated parameters, which reflects the sampling variability of the training data. Alternatively, predictions can be computed using each of the 200 bootstrap estimates, and their sample variance used to quantify prediction uncertainty. However, we note that in practice, prediction variance due to sampling variability in the training data is rarely reported. Users of published prediction models typically apply only the point estimates of model parameters, without accounting for the sampling variability of these estimates from the training data.

\subsection{\textbf{Prediction accuracy measures}}
\label{subsec:prediction.accuracy}

We follow the previous literature in choosing the prediction accuracy measures. For predicting competing risk outcomes, we use the time-dependent Brier score (BS) and the time-dependent AUC (Definition A in \citealt{Wu2018}), both designed for competing risk data and available in R package \texttt{tdROC}. For predicting the future longitudinal biomarkers, we use RMSE (root mean squared error). Since our data application is CKD research, we also use two commonly used measures in that field for predicting GFR, the P30 and P50 \citep{Levey1999}. They are the proportions of subjects for whom the predicted GFR falls within 30\% or 50\% of the observed GFR at the horizon. Higher AUC, lower BS, lower RMSE, higher P30 and P50 suggest better predictive accuracy. In the validation dataset, the GFR measurements may be irregularly spaced; when  there is no GFR data at the horizon, the closest measurement prior to the horizon is used.

\section{Data Application}
\label{sec:data_application}

We illustrate the proposed methodology by analyzing data from the African American Study of Kidney Disease and Hypertension (AASK). AASK is a longitudinal cohort study of 1,094 CKD subjects for up to 12 years. Detailed descriptions of AASK data were reported previously \citep{Li2016, Wu2020}. By the end of the follow-up, 318 subjects (29\%) reached ESRD and 176 subjects (16\%) died before reaching ESRD. We did not consider survival after ESRD in a semi-competing risk analysis because the patient experience and treatments were substantially different after reaching ESRD (\textbf{Remark 3.3}). We used two baseline predictors, age and gender, and three longitudinally measured predictive biomarkers: the estimated glomerular filtration rate (eGFR) using the MDRD equation, 24-hour urine protein to creatinine ratio (UPCr), and serum albumin. These biomarkers have known clinical relationships with the survival outcomes. 

\begin{figure}[p]
	\centering
	\includegraphics[width = 1\textwidth]{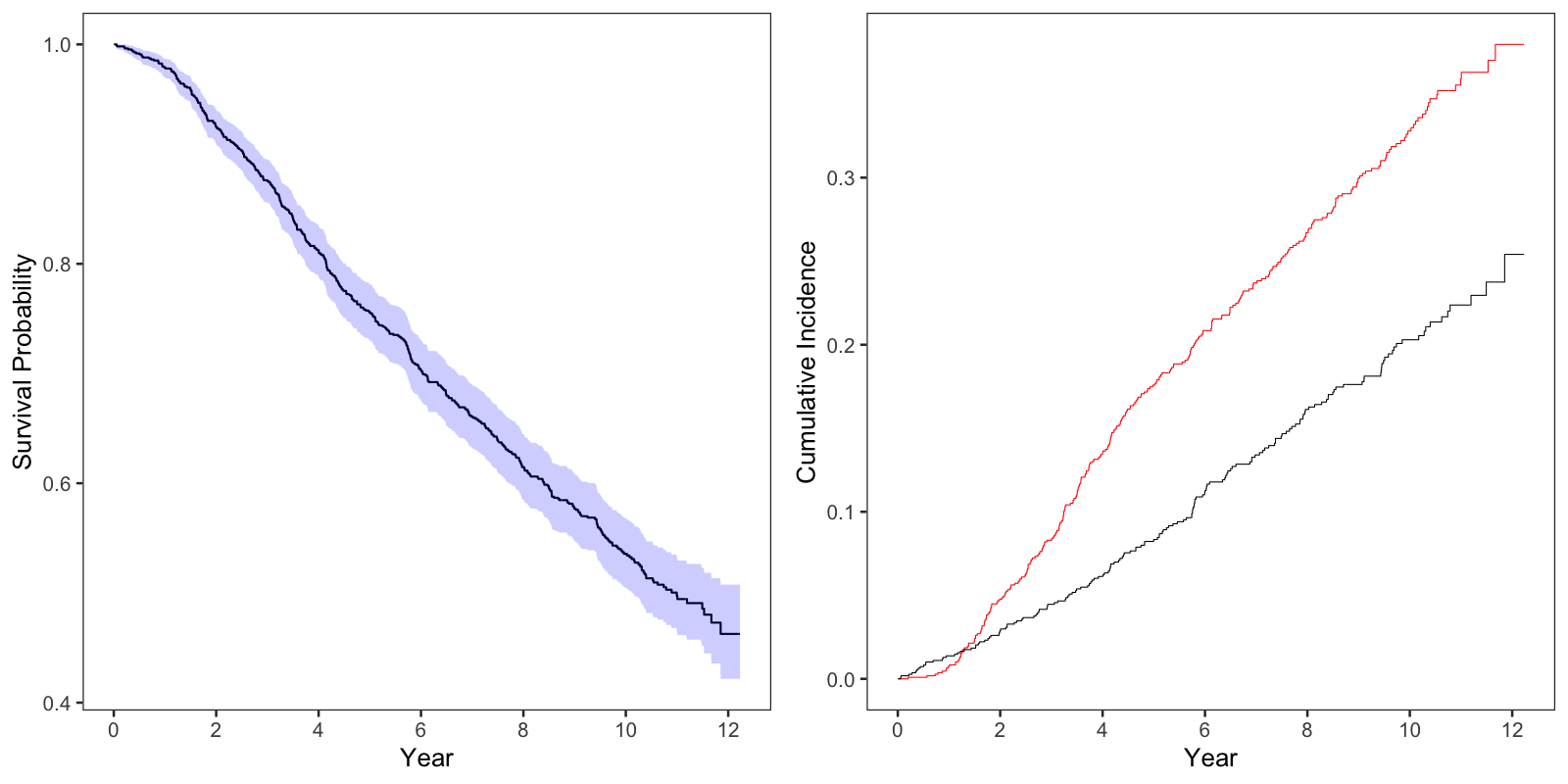}
	\caption{The Kaplan-Meier survival curve of the composite event of ESRD and death in the AASK study on the left and the cumulative incidence functions of ESRD (red) and death (black) on the right. }  \label{KM_CIF_AASK}
\end{figure}

\begin{figure}[h]
	\centering
	\includegraphics[width = 0.75\textwidth]{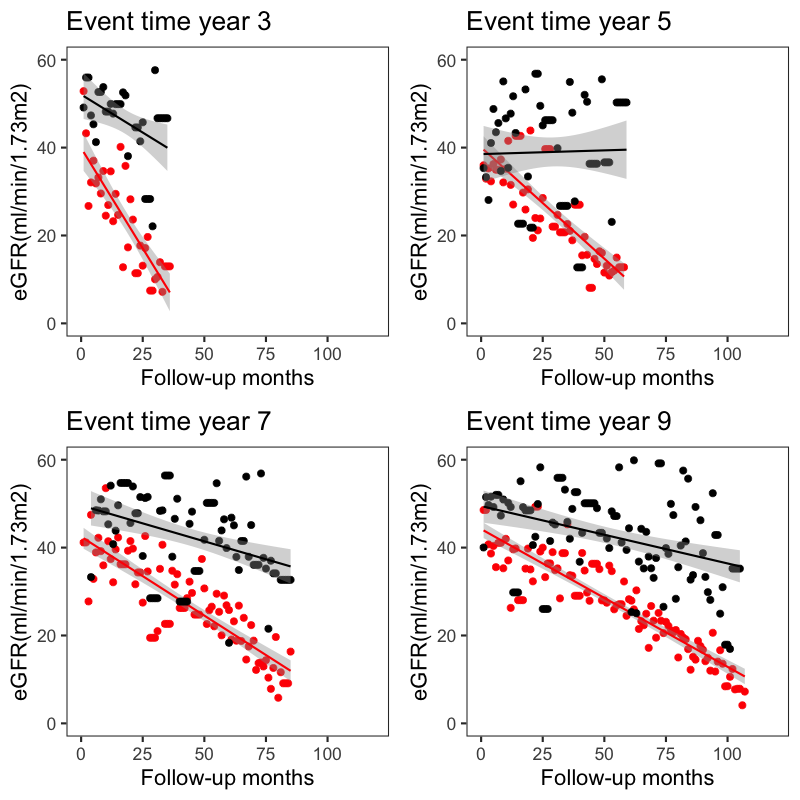}
	\caption{ \textit{The conditional mean trajectory (CMT) of eGFR at 4 selected composite event times in the AASK study. In each plot, we aggregated the longitudinal biomarker data from all subjects who had ESRD (red) or death (black) in that year. The dots show the mean biomarker by month, averaged across all subjects who had a measurement at that month. When the measurement times are non-informative, these biomarker means are unbiased estimators of the CMT at each month. A linear model is fitted to the monthly mean biomarker data for those with ESRD and death separately, with pointwise confidence intervals shown as gray bands. } } \label{cmt_egfr}
\end{figure}

Figure \ref{KM_CIF_AASK} displays the Kaplan-Meier curve of the composite endpoint of ESRD or death, whichever occurs first, and the cumulative incidence functions of ESRD and death separately. The distribution of the composite event in this population can be estimated up to 12 years, with more than 40\% of the events expected beyond that time. This supported the use of crBJM-TP. For comparison purposes, we also presented results from crBJM-EX. We set $\tau_{max} = 12$ years, the maximum observed event time in the data. We experimented with both vertical modeling and cause-specific hazard modeling as the survival sub-model, and they produced similar prediction accuracy. The reported results here came from crBJM with vertical modeling. Figure \ref{cmt_egfr} displays the CMTs of eGFR, which are approximately linear and suggest a clear association with the time of the composite endpoint and event types, with an interaction between the two. 

Including more predictive longitudinal biomarkers improves the prediction accuracy. To illustrate the benefit of using more longitudinal predictors, we considered crBJMs with one (crBJM-1, with eGFR), two (crBJM-2, with eGFR and UPCr), or three (crBJM-3, with eGFR, UPCr, and albumin) longitudinal biomarkers. With crBJM-3, the extrapolation model (EX) has 48 parameters and the two-part model (TP) has 72 parameters. The crBJM remains computationally convenient in this situation. Our EM algorithm achieved 100\% convergence rates in these analyses. Since this paper is mainly about prediction, we focus on prediction results here, and defer the extensive details on model specification, parameter estimation and interpretation, and sensitivity analysis to the Online Appendix. 

% The crBJM includes a competing risk regression model and a linear mixed model for each longitudinal parameter. These are conventional statistical models in which all the parameters are well identified. The number of parameters characterizing the correlation of multiple longitudinal biomarkers increases quadratically with the number of biomarkers, but these parameters are inevitable in any joint models with multivariate longitudinal biomarkers. The TP model has 50\% more parameters because it uses two parts to improve flexibility without extrapolation, but all the parameters in its LTS sub-model are identified.    

We used five-fold cross-validation to evaluate the prediction accuracy at a horizon of 3 years (Table \ref{AASK_DP}). Additional results under 1-year horizon are reported in the Online Appendix. As for the prediction of ESRD and death, Table \ref{AASK_DP} shows that crBJMs with more longitudinal biomarkers generally performed better than those with less, with crBJM-3-TP performing similarly to or better than all other methods in both AUC and BS. The TP methods generally performed better than the EX methods, likely because it is less prone to model misspecification when the tail distribution of the composite event cannot be estimated. The performance of crBJM-3-EX rose on par with their counterpart TP model (crBJM-3-TP), while crBJM-2-EX and crBJM-1-EX did not. The performance of EX models is more difficult to evaluate, probably because the accuracy of the extrapolation is unknown and cannot be verified from data. Nonetheless, crBJM-3-EX achieved similar predictive accuracy for ESRD and death as crBJM-3-TP did but with fewer parameters, making it an empirically useful model for prediction purposes. In predicting the eGFR, TP methods substantially outperformed EX methods. The BS appears to be more sensitive to between-method differences than the AUC, possibly because the latter is rank-based. The incidence of ESRD can be predicted with very high AUC because the treatment decision of dialysis or renal transplantation is strongly influenced by the patient's eGFR.

For comparison purposes, we fitted the SJM using the \texttt{JMBayes2} package in R. The model included all three longitudinal biomarkers (i.e., denoted by SJM-3 in Table \ref{AASK_DP}), whose trajectories were modeled by linear mixed models with subject-specific random intercepts and random slopes, and these random effects were shared with the cause-specific hazard models as covariates. The crBJMs outperformed the SJM in predicting ESRD and death. The results of eGFR prediction are mixed. However, it appears that the previous literature on SJM does not account for the absence of the composite event when predicting eGFR, which complicates the interpretation of this comparison. The crBJM is computationally much faster than the SJM. To complete the five-fold cross-validation on a personal computer with a 2.9 GHz CPU and 32 GB RAM, SJM-3 took 3 hours, whereas crBJM-3-TP with EM algorithm took 13 minutes. The CCA of crBJM can implemented with standard statistical software and the computing time is almost negligible. The computing time will multiply if the bootstrap is used for variance estimation. However, the bootstrap can be easily parallelized in a computer cluster, substantially reducing the computing time. Additionally, variance estimation is not essential in prediction problems (Section \ref{subsec:variance.estimation}). 

% However, it seems that the SJM does not condition on the absence of the composite event when predicting eGFR (Section 7.2 of \citealp{Rizopoulos2012}), which makes this comparison difficult. 

\begin{figure}[p]
	\centering
	\includegraphics[width=1.0\textwidth]{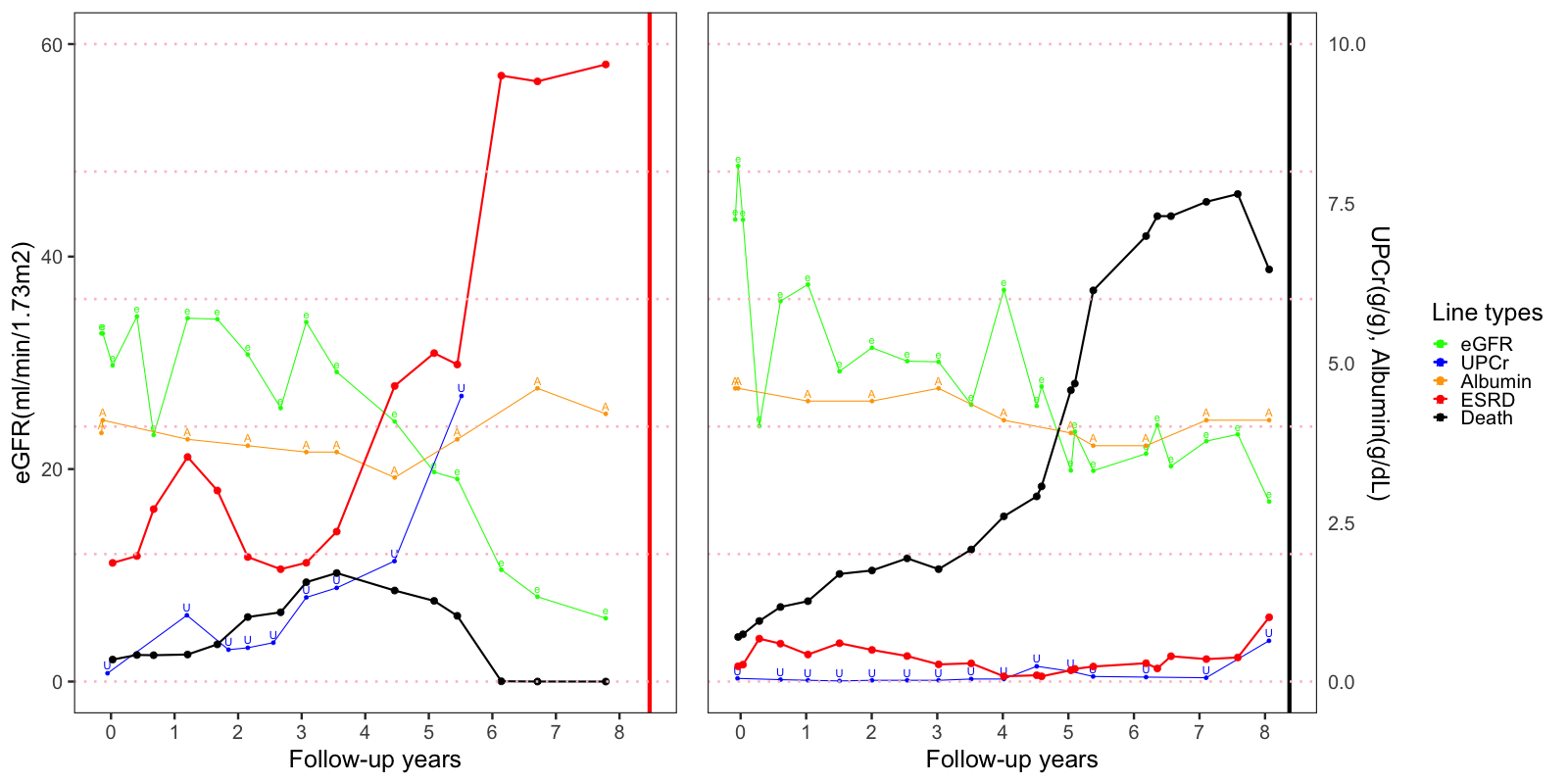}
	\caption{ \textit{Illustrating the dynamic prediction of ESRD and death with two AASK patients (patient A on the left and patient B on the right). The horizontal axis is the year since baseline. The longitudinal measurements of eGFR (green dots; connected by segments to form a continuous trajectory curve) are drawn to the vertical axis on the left; the longitudinal measurements of UPCr (blue dots and curve) and albumin (orange dots and curve) are drawn to the vertical axis on the right. The six horizontal dotted lines in the background mark the probability of 0, 0.2, 0.4, 0.6, 0.8, and 1.0, from bottom to the top. The predicted probabilities of ESRD (red dots; connected by linear segments to form a continuous curve) and death (black dots and curve) within a horizon of 3 years can be read from their positions among these six lines. The vertical red line marks the ESRD time of patient A; the vertical black line marks the death time of patient B. This dynamic prediction plot visualizes how the personalized, real-time risks of ESRD and death vary with the three longitudinal biomarkers. The prediction also incorporated baseline predictors.} } \label{AASK_individual_risk}
\end{figure}

\begin{figure}[p]
	\centering
	\includegraphics[width=0.9\textwidth]{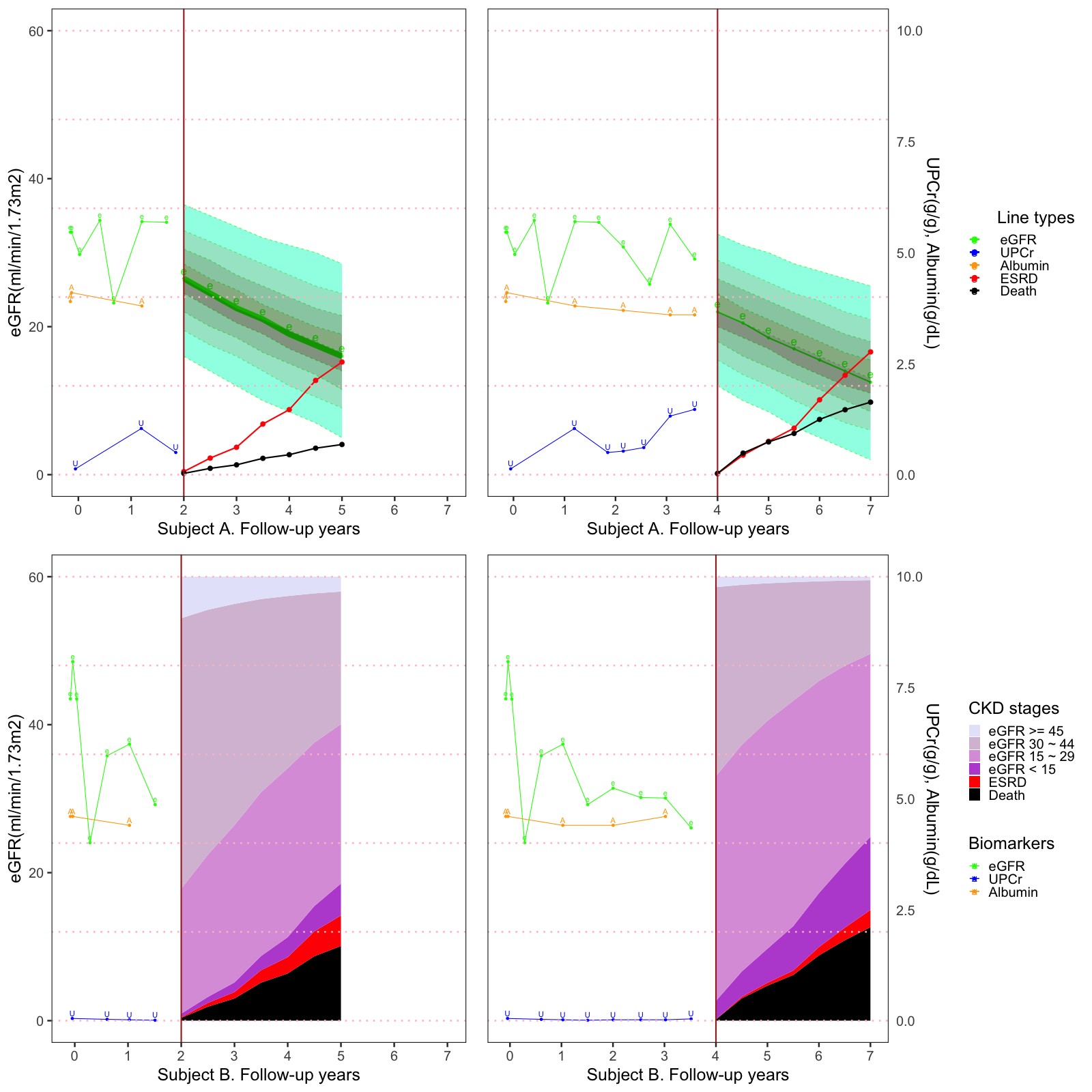}
	\caption{\textit{Illustrating the joint prediction of the future ESRD, death and eGFR distributions in two graphical presentations. The top plots show the prediction of AASK patient A at two prediction times (vertical brown line), based on the accumulating biomarker history to the left. The bottom plots show patient B. These are the same two patients in Figure \ref{AASK_individual_risk} and the plotting symbols are similar. The horizon is 3 years. With patient A, we plotted the predicted probabilities at 8 equally spaced future time points ($t = 0, 0.5, 1.0, ..., 3.0$). At each time, we calculated the density function of eGFR and its 9 quantiles (10\%, ..., 90\%). These quantiles are plotted to the vertical axis on the left. For better visualization, we colored the areas between adjacent quantile curves into 8 green shaded areas. The thick green curve in the middle indicates the predicted mode of eGFR. With patient B, we plotted the multistate representation of all predicted outcomes \citep{Hu2012}. The probabilities of death, ESRD and CKD stages (defined by eGFR; see Section 1) were calculated up to the horizon. These probabilities were drawn to the dotted horizontal lines in the background, and they sum up to 1 at each time point.} } \label{AASK_eGFR_quantiles_CKD}
\end{figure}

Figure \ref{AASK_individual_risk} depicts individual dynamic predictions of ESRD and death for two AASK patients. The patient on the left (A) had stable eGFR before year 3 but went into a steady decline from year 3 to year 8. The UPCr also increased over time, indicating increasing kidney damage. The predicted future probability of ESRD increases sharply in response. This patient eventually reached ESRD. The patient on the right had stable biomarkers and the predicted ESRD risk is low. The mortality risk, however, increases over time, presumably due to both aging and low ESRD risk (a competing risk phenomenon). Figure \ref{AASK_eGFR_quantiles_CKD} illustrates two useful graphical presentations of the joint prediction of competing risk and longitudinal outcomes. The first  uses the quantile curves to visualize how the predicted eGFR distribution conditional on absence of ESRD or death varies over time, in addition to the predicted cumulative incidence functions of ESRD and death. This patient's eGFR was within a stable range between the two prediction times, and the predicted ESRD was similar as a result. However, we can observe a subtle decrease in the predicted eGFR in the second prediction, perhaps due to the lower observed eGFR just prior to the prediction and increased UPCr. The cumulative incidence of death increases due to time elapsing and aging. The second graphical presentation is clinically more interpretable \citep{Hu2012}, where the future probabilities of being in each CKD stage and death are visualized. From year 2 to year 4, the eGFR stabilized around 30, which triggered the model to lower the probability of ESRD but enlarge the probability area corresponding to CKD stage 4 (eGFR between 15 and 29). The probability of death increased slightly due to time elapsing and aging.   

We performed sensitivity analyses to study whether the prediction accuracy is affected by the specifications for random effects in (\ref{BJM_model}), random intercept only or both random intercept and slope, and the choices of $\phi(.)$ in that model, $\phi(t) = \log(t)$ or $\phi(t)=t$. The random slope adds flexibility to the individual-level longitudinal trajectories within each survival stratum (i.e., given $\tT$) but also adds more parameters to the variance matrix for the random effects. A log transformation may fit the data better when $\tT$ is skewed. Table S4 of the Online Appendix shows that the prediction accuracy is not sensitive to these alternative formulations. We speculate that the survival time in the AASK data is not very skewed. For this reason, we used the simpler model with random intercept and without log transformation.

\section{Simulation}
\label{sec:Simulation}

We conducted simulations to study the finite sample performance of the proposed estimation procedure for both TP and EX models, and to compare the relative efficiency of CCA vs. EM algorithm. The CCA and EM methods both produce consistent estimation, but CCA is easier to use, making it worthwhile to assess its cost on efficiency. Since the crBJM-EX and crBJM-TP have different model specifications, we conducted separate simulation studies for them. In each study, we simulated 300 independent Monte Carlo datasets of $n = 300$ or $1,000$ subjects, analyzed each dataset, and calculated the bias and the relative efficiency. We simulated one baseline covariate and three longitudinal biomarkers. The longitudinal sub-models had a linear CMT and a random intercept. The competing risk model is a vertical model with a Weibull regression for the time to composite event. The censoring rate is 40\% in the crBJM-EX data. For crBJM-TP, the overall censoring rate is 40\% and 20\% subjects were censored at $\tau_{max}$. 

% We already compared the predictive performance of crBJM and SJM in the AASK data. We do not compare them in the simulations because they operate under different modeling assumptions and it is impossible to simulate data so that both are correctly specified. Within the likelihood framework, consistent and efficient estimation of the crBJM guarantees optimal predictive accuracy. 

% The crBJM-EX and crBJM-TP were fit to their respective datasets so that we can evaluate the estimation methods under the correct model specification. The CCA was used to obtain the initial values for the EM. 

\begin{figure}[p]
	\centering
	\includegraphics[width = 1.0\textwidth]{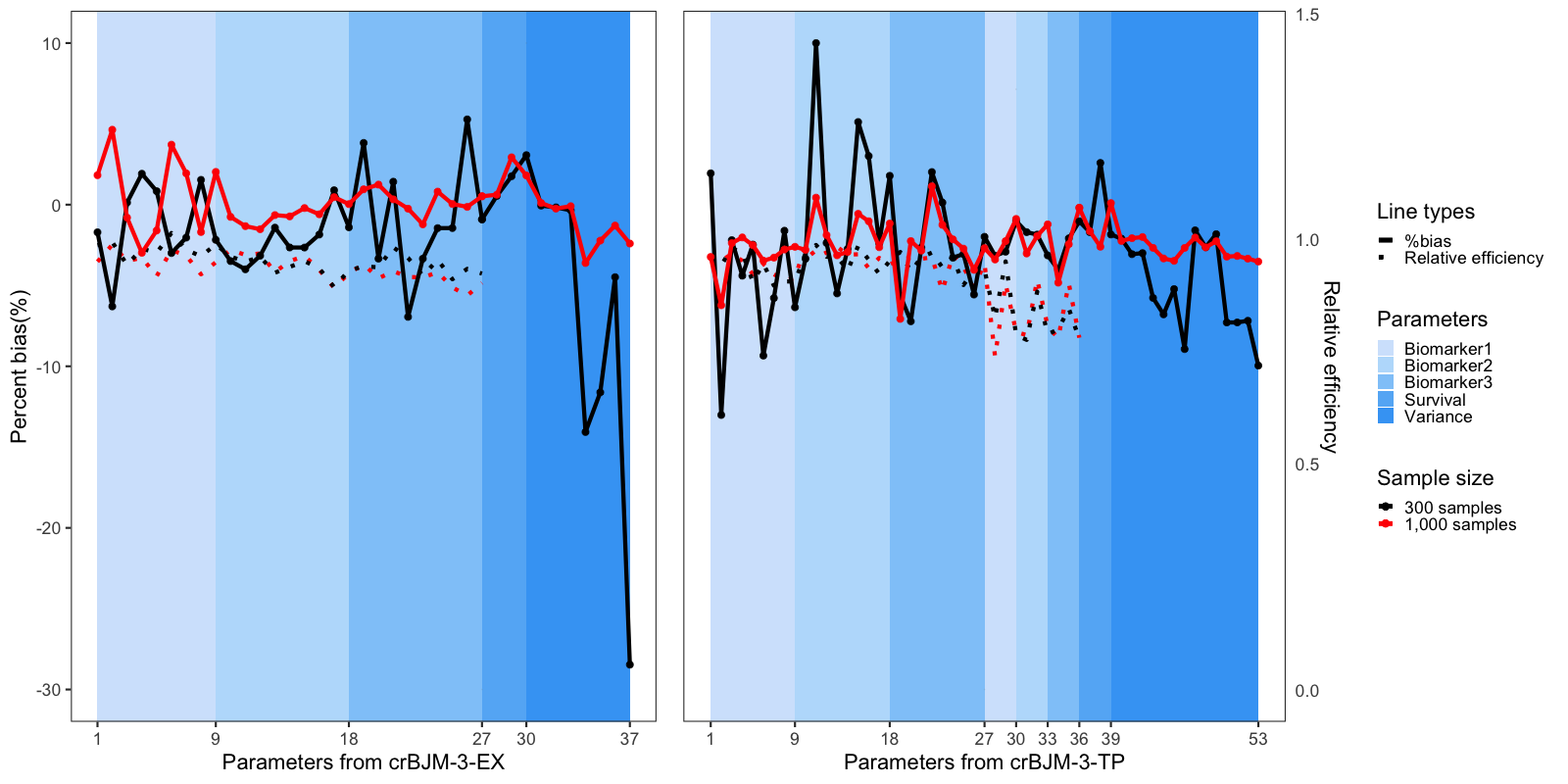}
	\caption{ \textit{Bias and efficiency of the crBJM estimation in simulations (left: crBJM-3-EX; right: crBJM-3-TP). The horizontal axis is the index of parameters, grouped by their roles in the model. For better visualization, we used different background colors for different groups of parameters. 
			For the EX model, the parameter groups from left to the right are longitudinal sub-models for biomarkers 1-3, survival sub-model, and variance-covariance matrix. For the TP model, the parameter groups from left to the right are longitudinal sub-models for biomarkers 1-3, longitudinal LTS sub-models for biomarkers 1-3, survival sub-model, and variance-covariance matrix. The number of parameters in each parameter group can be read from the horizontal axis. The 4 curves demonstrate the percent bias (drawn to the vertical axis on the left) and relative efficiency (drawn to the vertical axis on the right) of each estimated parameter from the EM algorithm, under the sample size of $n=300$ or $n=1,000$. The relative efficiency is measured by the ratio of the empirical standard deviations of the point estimators from the EM and CCA. Values less than 1 indicate that EM is more efficient than CCA; smaller values indicate more efficiency improvement.} }  \label{Sim_consistency_efficiency}
\end{figure}

All EM algorithms converged under the criterion that all longitudinal sub-model parameters differed by no more than $10^{-4}$ in consecutive iterations. Further details of the data-generating schemes are in the Online Appendix. Figure \ref{Sim_consistency_efficiency} shows that the finite-sample biases of the EM estimators are generally small and drop to nearly zero when the sample size increases from $300$ to $1,000$, supporting consistent estimation. The EM algorithm exhibits notable efficiency improvements over the CCA, especially in the LTS sub-model parameters, likely because the CCA only utilizes subjects at risk by $\tau_{max}$ to estimate the LTS, which can result in a small sample size.

\section{Discussion}
\label{sec:Discussion}

This paper extends the BJM methodology originally proposed by \citet{Shen2021} for dynamic prediction, with several novel developments. These include joint prediction of survival and longitudinal outcomes, models for competing risks, two formulations of the BJM (extrapolation and two-part), links to other joint modeling frameworks, and a theoretical comparison between static and dynamic prediction. To provide a comprehensive overview, we begin with the core modeling concepts --- emphasizing likelihood factorization, model assumptions, and related literature --- before addressing extensions for competing risks and estimation strategies. R code implementing the BJM is available at \url{https://github.com/liwh0904/BJM}.

Of the two dynamic prediction approaches (Section \ref{subsec:overview.DP}), joint modeling offers several advantages over landmark modeling, including efficient likelihood-based estimation and natural incorporation of longitudinal history. Landmark modeling, however, is easier to implement using standard software and can handle a large number of predictors with manageable computation \citep{ferrer2019individual, Li2023}. The proposed BJM reduces computational burden and implementation complexity to a level comparable to landmark modeling. Moreover, to our knowledge, no landmark approach can perform joint dynamic prediction of longitudinal and survival outcomes as defined in Section \ref{subsec:overview.DP}. 

Recent developments in dynamic prediction using joint modeling have trended toward incorporating a growing number of longitudinal predictors. However, the well-known computational challenge of the SJM poses a significant barrier \citep{Rizopoulos2012, Elashoff2016}. Some studies address this by ``brute-force'' with advanced computation techniques \citep{JMBayes2, mauff2020joint, Rustand2023, LavalleyMorelle2023}. Noting from (\ref{DP_risk}) and (\ref{DP_eGFR}) that any well-specified joint model for the longitudinal and survival data can support dynamic prediction, we sought an alternative route to overcome this challenge with novel model formulation, which led to the development of the BJM.    

The joint dynamic prediction formulas (\ref{DP_risk}) and (\ref{DP_eGFR}), as well as the likelihood factorization (\ref{eq:factorize}), are mathematical identities. Thus, optimal prediction accuracy is expected when each sub-model in the factorization is ``correctly'' specified. In practice, this requires that the sub-models closely approximate the data while guarding against under- and overfitting. A current limitation of this paper is the use of parametric longitudinal sub-models. However, model checking is straightforward within the BJM framework using consistent CCA-based estimation. Future work is needed to incorporate nonparametric specifications, enhancing modeling flexibility and mitigating model misspecification. However, the basic modeling framework outlined in this paper remains the same. 

The BJM framework can also be extended to handle a mixture of continuous and categorical longitudinal variables by replacing the multivariate linear mixed model with a multivariate generalized linear mixed model (GLMM). However, since the GLMM usually has intractable integrals in its likelihood, the resulting BJM loses the advantage of having only one-dimensional numerical integration. This challenge is addressed in a separate paper \citep{WenhaoPaper3.arxiv}.

% At a chosen $\tau_{max}$, the percent of subjects contributing to the LTS estimation is $P( \tT > \tau_{max}) P( C > \tau_{max} )$. The censoring distribution may depend on staggered study entry, administrative censoring, and dropout patterns. When $P( C > \tau_{max} )$ is low, we may need to decrease $\tau_{max}$ to ensure enough sample size for the LTS estimation. The longitudinal and survival data beyond $\tau_{max}$ are artificially censored as a result. 

\section*{Conflict of interest}
The authors declare that they have no conflict of interest.

\section*{Supplement}

The Online Appendix is available for download from the journal's website.

%\bibliographystyle{vancouver}
%\bibliography{MBJMRef}

\bibliographystyle{spbasic}
\bibliography{BJMcmprskRefLDA.bib}

\newpage

\begin{table}[p]
	\centering
	\caption{Prediction accuracy results of crBJM and SJM in the AASK data analysis. The prediction accuracy was evaluated among at-risk subjects at Year 1-4 after baseline. The prediction horizon is 3 years. SJM: shared random effects joint model. }\label{AASK_DP}
	\begin{tabular}{ccccccccccccccc}
		\hline
		Accuracy & Type& Model&  Year1 & Year2 & Year3 & Year4   \\ 
		\hline
		AUC  &ESRD& crBJM-3-TP   &0.9068 & 0.9378 & 0.9267 & 0.9008  \\
		AUC  & ESRD& crBJM-2-TP    &0.9068 & 0.9358 & 0.9259 & 0.8987  \\ 
		AUC  &  ESRD&crBJM-1-TP  &0.8727 & 0.9135 & 0.9322 & 0.9135   \\ 
		AUC  &ESRD& crBJM-3-EX   &    0.9057 & 0.9398 & 0.9342 & 0.9182  \\ 
		AUC  &ESRD& crBJM-2-EX    & 0.9096 & 0.9492 & 0.9527 & 0.9587  \\ 
		AUC  & ESRD&crBJM-1-EX    &0.8796 & 0.9314 & 0.9358 & 0.9277  \\
		AUC  & ESRD&SJM-3    &  0.8941 & 0.8945 & 0.9506 & 0.9449  \\ 
		
		AUC  & Death&  	 crBJM-3-TP   & 0.7354 & 0.6430 & 0.6736 & 0.6725  \\ 
		AUC  & Death&   crBJM-2-TP	&0.7083 & 0.6099 & 0.6569 & 0.6620\\ 
		AUC  & Death&  	 crBJM-1-TP & 0.6586 & 0.6387 & 0.6162 & 0.5888  \\
		AUC  & Death& 	 crBJM-3-EX   &  	  0.5999 & 0.6554 & 0.6403 & 0.6295  \\
		AUC  & Death& 	 crBJM-2-EX   &  	0.5873 & 0.6114 & 0.6098 & 0.6163  \\ 
		AUC  & Death& 	 crBJM-1-EX   &  	  0.6242 & 0.6497 & 0.6605 & 0.6566  \\
		AUC  & Death& 	 SJM-3  &  	  0.6709 & 0.6316 & 0.6243 & 0.6874   \\ 
		\hline
		
		BS&ESRD & crBJM-3-TP      & 0.0672 & 0.0585 & 0.0677 & 0.0714   \\ 
		BS&ESRD &  crBJM-2-TP      &0.0659 & 0.0586 & 0.0677 & 0.0710  \\ 
		BS&ESRD &  crBJM-1-TP      &0.0776 & 0.0690 & 0.0652 & 0.0672   \\ 
		BS&ESRD &  crBJM-3-EX      &0.0686 & 0.0582 & 0.0673 & 0.0693 \\ 
		BS&ESRD &  crBJM-2-EX       & 0.0677 & 0.0640 & 0.0739 & 0.0809  \\ 
		BS&ESRD &  crBJM-1-EX       & 0.0809 & 0.0786 & 0.0812 & 0.0921  \\ 
		BS&ESRD &  SJM-3      &0.1177 & 0.1151 & 0.1078 & 0.0718 \\ 					
		
		BS&Death&   crBJM-3-TP      & 0.0422 & 0.0512 & 0.0642 & 0.0748 \\ 
		BS&Death&   crBJM-2-TP      &0.0425 & 0.0514 & 0.0646 & 0.0741  \\ 
		BS&Death&   crBJM-1-TP      &0.0422 & 0.0486 & 0.0588 & 0.0674 \\ 
		BS&Death&  crBJM-3-EX     &0.0442 & 0.0504 & 0.0614 & 0.0710   \\ 
		BS&Death&  crBJM-2-EX      &0.0550 & 0.0699 & 0.0878 & 0.1020 \\
		BS&Death&  crBJM-1-EX      &0.0547 & 0.0694 & 0.0868 & 0.1009  \\ 
		BS&Death&  SJM-3    &0.0867 & 0.1081 & 0.1279 & 0.1306 \\  
		\hline		
		RMSE &eGFR&  crBJM-3-TP   &8.1559 & 8.7276 & 8.4907 & 8.2903  \\
		RMSE &eGFR&  crBJM-3-EX & 10.6981 & 10.6853 & 9.5622 & 8.4871    \\
		RMSE &eGFR&  SJM-3    & 8.8914 &8.9091& 8.0943& 6.8213  \\
		
		P30 &eGFR&  crBJM-3-TP  &0.9156 & 0.9241 & 0.9469 & 0.9308\\
		P30& eGFR&  crBJM-3-EX  &0.7891 & 0.8254 & 0.8607 & 0.9000  \\
		P30& eGFR&   SJM-3      &0.9213 &0.9195 &0.9110& 0.9466\\
		
		P50 &eGFR&  crBJM-3-TP   &0.9600 & 0.9709 & 0.9812 & 0.9853  \\
		P50&eGFR &  crBJM-3-EX &0.9553 & 0.9520 & 0.9614 & 0.9742  \\
		P50 &eGFR&  SJM-3   &0.9683 &0.9831 &0.9874 &0.9871\\
		\hline
		\multicolumn{3}{c}{\# at-risk patients in AASK data}&1065&1005&955&875\\
		\hline
	\end{tabular}
	
\end{table}

\end{document}